\documentstyle[a4,12pt]{article}

\newcommand{\nit}{\noindent}
\newcommand{\nl}{\newline}

\newcommand{\dsp}{\displaystyle} 
\newcommand{\vs}[1]{\vspace{#1 ex}}
\newcommand{\bfr}{\begin{flushright}}
\newcommand{\efr}{\end{flushright}}
\newcommand{\bc}{\begin{center}}
\newcommand{\ec}{\end{center}} 
\newcommand{\ben}{\begin{enumerate}}
\newcommand{\een}{\end{enumerate}} 

\newcommand{\be}{\begin{equation}}
\newcommand{\ee}{\end{equation}}
\newcommand{\ba}{\begin{array}}
\newcommand{\ea}{\end{array}}
\newcommand{\ct}{\cite}
\newcommand{\bit}{\bibitem}

\newcommand{\ag}{\alpha}

\newcommand{\gam}{\gamma}
\newcommand{\del}{\delta}

\newcommand{\thg}{\theta}
\newcommand{\kg}{\kappa}
\newcommand{\lb}{\lambda}

\newcommand{\rg}{\rho}
\newcommand{\fg}{\phi}

\newcommand{\og}{\omega}
\newcommand{\Gam}{\Gamma}

\newcommand{\Lb}{\Lambda}

\newcommand{\lh}{\left(}
\newcommand{\rh}{\right)}

\newcommand{\pl}{\partial}

\newcommand{\kgp}{\kg_+} 
\newcommand{\kgt}{\kg_{\times}}
\newcommand{\dbv}{\del {\bf v}}

\begin{document}

\begin{flushright} 
NIKHEF/99-019 

\end{flushright}

\begin{center} 
{\Large{\bf Cyclotron motion in a gravitational-wave}}\\
\vs{3} 

{\Large{\bf background}} \\
\vs{5}

{\large J.W.\ van Holten} \\
\vs{3}

{\large NIKHEF, Amsterdam NL}\\ 
\vs{3}

{\tt t32@nikhef.nl} \\
\end{center} 
\vs{7}

\nit
{\small {\bf Abstract}}
{\footnotesize 
We examine the motion of a relativistic charged particle in 
a constant magnetic field perturbed by gravitational waves 
incident along the direction of the magnetic field. We 
apply a generalized energy conservation law to compute the 
variations of the kinetic energy of the particle during 
the passage of the waves. We also explicitly compute the 
change in the orbit due to a wave with constant curvature.}
\vs{2} 

\nit
{\bf Introduction} \nl 
It is well-known that the Einstein equations of general 
relativity admit wave-solutions. Indeed, approximate as well 
as exact solutions were found early on by Einstein, Brinkmann 
and others; for reviews and references to the early literature, 
see e.g.\ \ct{Wein}-\ct{jw1}. The relevance of these waves 
to astrophysics has been beautifully demonstrated by the 
observations of binary pulsars, in particular the pulsar 
PSR 1913 +16 discovered by Hulse and Taylor \ct{HT}, 
which loses energy precisely at the rate predicted by 
general relativity \ct{TW}. 

As the interaction of gravitational waves with matter is
extremely weak, very sensitive instruments will be 
needed to detect them. Bar antennae and laser interferometers 
are the most advanced instruments developed so far. It 
is expected that in the near future they will be able to 
search for gravitational waves with a strain sensitivity of 
the order of $h \sim 10^{-21}/\sqrt{\mbox{Hz}}$ or better in 
the frequency range between 100 -- 1000 Hz \ct{KT}. 

Recently the question has come up, whether it is feasible to 
search for gravitational waves using particles accelerators
\ct{ZZ}. Measurements of the ground motion, including tidal 
motion induced by the moon, have been made with the LEP 
accelerator at CERN with great precision \ct{cern}. In order 
to answer that question, in this paper we study a charged 
particle orbiting in a constant magnetic field under the 
influence of gravitational waves. Although this is a rather 
simplified model for a particle accelerator, it is quite 
useful to obtain estimates of orders of magnitude. We further 
simplify the treatment by restricting ourselves to waves 
incident perpendicular to the plane of the cyclotron orbit,
which guarantees maximal sensitivity of the particle to the 
gravitational perturbations, and by neglecting the effects 
of synchrotron radiation which causes the particle to lose 
energy during its motion. However, our model has the 
advantage that it allows a fully relativistic treatment, 
and that it is a priori not restricted to a weak-field 
approximation. Our calculations indicate, that particle 
accelerators are not useful as gravitational wave detectors 
in the frequency range of the bar antennae and laser 
interferometers. However, it is open to discussion whether 
they might become useful in the range of very high frequencies, 
where a cosmic background ---in the simplest case a thermal 
background with a temperature of the order of 1 K--- might 
show itself. 
\vs{2} 

\nit
{\bf Equations of motion} \nl 
The equation of motion for a relativistic charged 
particle in gravitational and electro-magnetic 
background fields is 
\be 
\frac{D^2 x^{\mu}}{D\tau^2}\, =\, 
\ddot{x}^{\mu}\, +\, \Gam_{\lb\nu}^{\;\;\;\;\mu} 
  \dot{x}^{\lb} \dot{x}^{\nu}\, =\, \frac{q}{m}\, 
  F^{\mu}_{\;\;\nu}\, \dot{x}^{\nu}. 
\label{2.1}
\ee 
Here $\Gam_{\lb\nu}^{\;\;\;\;\mu}$ is the Riemann-Christoffel 
connection, $F_{\mu\nu}$ is the electro-magnetic field-strength 
tensor, and the overdot denotes differentiation w.r.t.\ proper 
time $\tau$, which is related to the invariant co-ordinate 
interval along the worldline by 
\be 
c^2 d\tau^2\, =\, - g_{\mu\nu}\, dx^{\mu} dx^{\nu}. 
\label{2.2}
\ee 
This equation actually expresses the existence of a constant 
of motion
\be
H\, =\, g_{\mu\nu}\, \dot{x}^{\mu} \dot{x}^{\nu} = -c^2.
\label{2.3}
\ee
Indeed, differentiation of $H$ w.r.t.\ proper time and 
using the equation of motion while taking into account the 
anti-symmetry of $F_{\mu\nu}$ shows that 
\be 
\dot{H}\, =\, 0, 
\label{2.4}
\ee 
as required by the invariance of the speed of light. 

Gravitational waves propagating in a fixed direction, say 
the $+z$-axis, can be described by metrics of the form 
\ct{MTW,KSMH,jw1}
\be
g_{\mu\nu}\, dx^{\mu} dx^{\nu}\, =\, 
  -\, dudv\, -\, K(u,x,y) du^2\, +\, dx^2\, +\, dy^2.
\label{2.5}
\ee 
Here $x$ and $y$ are cartesian co-ordinates in the 
transverse plane, and $u$ and $v$ are the light-cone 
co-ordinates defined by 
\be
u\, =\, ct\, -\, z, \hspace{3em} v\, =\, ct\, +\, z. 
\label{2.6}
\ee 
Clearly, the metric component $K$ represents the deviation 
of the metric from flat minkowski space-time. Therefore all 
gravitational effects are determined by the behaviour of this 
quantity. This can be seen from the curvature components; in 
particular the only non-trivial component of the Einstein and 
Ricci curvature tensors is:
\be
G_{uu}\, =\, R_{uu}\, =\, - \frac{1}{2}\, \lh \pl_x^{\,2} + 
  \pl_y^{\,2} \rh\, K, 
\label{2.7}
\ee 
all other components vanishing identically. Therefore the 
Einstein equations in empty space are satisfied if $K$ 
is a harmonic function in the $x$-$y$-plane: 
\be 
\lh \pl_x^{\,2} + \pl_y^{\,2} \rh\, K\, =\, 0.
\label{2.8}
\ee 
Now constant solutions, and solutions linear in $(x, y)$, 
give not only a vanishing Ricci curvature, but a completely 
vanishing Riemann tensor; therefore they describe flat 
minkowski space-time in non-standard co-ordinates, and we 
exclude these solutions here. The simplest non-trivial 
solution of eq.(\ref{2.8}) then is 
\be
K(u,x,y)\, =\, \kgp(u)\, (x^2 - y^2)\, +\, 
    2 \kgt (u)\, xy.
\label{2.9}
\ee
In plane polar co-ordinates $(\rg, \thg)$ this can be 
written as 
\be
K\, =\, \rg^2 \lh  \kgp \cos 2 \thg + \kgt \sin 2 \thg \rh, 
\label{2.10}
\ee 
showing that the solutions have the periodicity $\pi$ 
characteristic of quadrupole waves. We also observe,
that the only non-vanishing components of the Riemann 
curvature tensor are 
\be 
R_{uxux}\, =\, - R_{uyuy}\, =\, - \kgp, \hspace{2em} 
R_{uxuy}\, =\, R_{uyux}\, =\, - \kgt. 
\label{2.11}
\ee 
Thus the curvature components are constant throughout the 
transverse plane; such waves are called $pp$-waves in the 
literature. Finally, it should be mentioned that the 
curvature components equal those computed in the 
$TT$--co-ordinates in the usual weak-field approximation. 
However, no weak-field assumption is made here. 

Having specified the gravitational field assumed to be 
present, we next introduce a constant magentic field $B$ 
in the same direction as the waves, i.e.\ the $z$-direction.
Therefore the electro-magnetic field-strength tensor has 
non-zero components 
\be
F_{xy}\, =\, - F_{yx}\, =\, B, 
\label{2.12}
\ee 
whilst all other components vanish. It is now easy to work 
out the equations of motion \ct{jw1,jw2}: 
\be 
\ba{ll}
\ddot{u}\, =\, 0, & \\ 
 & \\
\dsp{ \ddot{x}\, =\, -\frac{1}{2}\, \dot{u}^2 K_{,x}\, +\, 
  \frac{qB}{m}\, \dot{y}, }& \dsp{ 
  \ddot{y}\, =\,  -\frac{1}{2}\, \dot{u}^2 K_y\, -\,  
  \frac{qB}{m}\, \dot{x}. }
\ea  
\label{2.13}
\ee 
We have deliberately left out the equation for $v$, as
we replace it by the conservation law (\ref{2.3}): 
\be 
\dot{u} \dot{v}\, +\, K \dot{u}^2\, =\, \dot{x}^2\, +\, 
  \dot{y}^2\, +\, c^2. 
\label{2.14}
\ee 
From the first equation (\ref{2.13}) we infer that 
$\dot{u}$ is itself a constant of motion:
\be
\dot{u}\, =\, c\, \frac{dt}{d\tau}\, \lh 1 - \frac{v_z}{c} \rh\, 
 =\, \gam c, 
\label{2.15}
\ee 
with $\gam$ a dimensionless constant and $v_z = dz/dt$. Using 
that 
\be
\dot{u} \dot{v}\, =\, c^2 \dot{t}^2 - \dot{z}^2, 
\label{2.16}
\ee 
we can then deduce from eq.(\ref{2.14}) that 
\be 
\frac{dt}{d\tau}\, =\, 
\sqrt{\dsp{\frac{1 - \gam^2 K}{1 - {\bf v}^2/c^2}}}, 
\label{2.17}
\ee 
where ${\bf v} = d{\bf r}/dt$ is the velocity 3-vector in the 
lab frame. Combining the results (\ref{2.15}), (\ref{2.17}) 
finally allows us to obtain an expression for the constant 
$\gam$: 
\be 
\frac{1}{\gam^2}\, =\, \frac{1 - {\bf v}^2/c^2}{(1 - v_z/c)^2}\, 
   +\, K. 
\label{2.18}
\ee 
We can interpret this result as a generalization of the 
usual energy conservation law, with $K$ representing the 
gravitational potential. In particular, in the absence of 
a gravitational field $(K = 0)$ and for motion in the 
$x$-$y$-plane $(v_z = 0)$ we find the well-known expression 
\be 
\gam\, =\, \frac{1}{\sqrt{1 - {\bf v}^2/c^2}}. 
\label{2.18.1}
\ee

Eqs.(\ref{2.15}) and (\ref{2.18}) present the general solution 
to the equations of motion in the $(u,v)$, or equivalently 
$(z,t)$, directions. It remains to solve the equations of motion 
in the tranverse plane. We first notice, that equation (\ref{2.15}) 
allows us to write $cd\tau = du/\gam$; then the tranverse 
equations of motion (\ref{2.13}) can be cast into the form 
\be 
\ba{lll} 
x^{\prime\prime}\, -\, k_c y^{\prime}\, +\, \kgp x\, 
 +\, \kgt y & = & 0, \\
 & & \\
y^{\prime\prime}\, +\, k_c x^{\prime}\, -\, \kgp y\, 
 +\, \kgt x & = & 0. 
\ea 
\label{2.19}
\ee 
Here the prime denotes a derivative w.r.t.\ $u$, and 
\be 
k_c \, =\, \frac{qB}{mc\gam}. 
\label{2.20}
\ee 
Clearly, the precise solutions depend on the specific form 
of the curvature components $\kgp(u)$ and $\kgt(u)$. 
\vs{2} 

\nit
{\bf Discussion} \nl 
In the absence of a gravitational field $(K = 0)$, the 
transverse equations of motion (\ref{2.19}) have well-known 
solutions describing circular cyclotron orbits: 
\be 
x(u)\, =\, a \cos k_c u\, +\, b \sin k_c u, \hspace{2em} 
y(u)\, =\, b \cos k_c u\, -\, a \sin k_c u,  
\label{3.1}
\ee 
or equivalently in polar co-ordinates
\be
\rg = \rg_0 = \sqrt{a^2 + b^2}\, =\, constant, \hspace{3em} 
\thg\, =\, \og_0 t\, =\, \frac{qBt}{m\gam}. 
\label{3.2}
\ee 
Under the assumption that $v_z = 0$, $\gam$ is given here 
by eqn.(\ref{2.18.1}). Then it is straightforward to show 
by expressing ${\bf v}^2$ in polar co-ordinates as $\rg^2 
\og_0^2$ that 
\be 
\gam^2\, =\, \frac{1}{1 - {\bf v}^2/c^2}\, =\, 1\, +\, 
  \rg^2 \lh \frac{qB}{mc} \rh^2. 
\label{3.3}
\ee 
Therefore, if $\gam \gg 1$, we have from eq.(\ref{2.20})
\be 
k_c\, =\, \frac{1}{\rg}. 
\label{3.3.1}
\ee
We now ask, how the cyclotron motion is affected by the 
passage of a gravitational wave. A first consequence 
follows from the conservation law (\ref{2.18}). Let 
${\bf v}_0$ denote the planar velocity in the absence of 
a gravitational field. Then 
\be 
\gam\, =\, \frac{1}{\sqrt{1 - {\bf v}_0^2/c^2}}. 
\label{3.4}
\ee 
Consider the relativistic case $\gam \gg 1$; as 
a gravitational wave passes the particle, its velocity is 
changed from ${\bf v}_0$ to ${\bf v} = {\bf v}_0 + \dbv$. 
We immediately derive that the change in the kinetic 
energy is given by
\be 
K\, =\, -\, \del \lh \frac{1 - (\bf v)^2/c^2}{(1 - v_z/c)^2} 
  \rh\, \approx\, 2\, \frac{{\bf v}_0 \cdot \dbv}{c^2}\, 
  -\, 2\, \frac{\del v_z}{\gam^2 c}.
\label{3.5}
\ee 
Note that as the initial velocity is purely tangential, 
the first term on the r.h.s.\ actually describes only the 
change in tangential velocity: 
\be 
{\bf v}_0 \cdot \dbv\, =\, \rg^2_0\, \og_0\, \del \og\, 
  =\, {\bf v}_0^2\, \frac{\del \og}{\og_0}. 
\label{3.6}
\ee 
The last term in (\ref{3.5}) we neglect because of the 
relativistic nature of the particle. Then we actually can 
write more concisely  
\be 
\del \lh \frac{{\bf v}^2}{c^2}\rh \, =\, 
   K\, \approx\, 2\, \frac{\del \og}{\og_0}. 
\label{3.7}
\ee 
Thus the change in total kinetic energy is simply given 
by the gravitational potential, the expression for which 
is that of eq.(\ref{2.10}):
\[
K\, =\, \rg^2\, (\kgp \cos 2\thg + \kgt \sin 2 \thg). 
\]
As the perturbation by the gravitational wave is inevitably 
small, we may to first approximation take $\rg$ and $\thg$ 
to be given by the unperturbed values: 
\be 
K\, \approx\, \rg_0^2\, (\kgp(u) \cos 2\og_0 t + \kgt(u)
   \sin 2 \og_0 t), 
\label{3.8}
\ee 
with in the same approximation $u = ct$, as $z \approx 0$. 

As an example, consider a monochromatic wave with $+$--polarization
of strain amplitude $h_+$ and angular frequency $\og_+ = 2\pi  f_+$.
The curvature components for this field are 
\be 
\kgp\, =\, \frac{\og_+^2}{2c^2}\, h_+\, \cos (\og_+ t -\ag), 
 \hspace{3em} \kgt\, =\, 0.
\label{3.9}
\ee 
Here $\ag$ is the phase difference at $t = 0$ between the 
gravitational wave and the cyclotron motion. It follows 
that 
\be 
\ba{lll}
K & = & \dsp{ \frac{\rg_0^2 \og_+^2}{2c^2}\, h_+\, 
 \cos 2\og_0 t\, \cos (\og_+t - \ag) }\\
 & & \\
 & = & \dsp{ \frac{\rg_0^2 \og_+^2}{4c^2}\, 
 h_+\, \left[ \cos \lh (\og_+ - 2\og_0)\,t - \ag \rh + 
 \cos \lh (\og_+ + 2\og_0)\,t - \ag \rh \right].} 
\ea 
\label{3.10}
\ee 
In terms of ordinary frequencies $f_+$ for an accelerator 
of the size of LEP, which has a radius $\rg_0$ = 4.2 km, this 
gives 
\be 
\del \lh \frac{{\bf v}^2}{c^2}\rh\, =\, K\, \leq\, 
 0.4 \times 10^{-8}\, \lh \frac{f_+}{1\, \mbox{Hz}}\rh^2\, h_+. 
\label{3.11}
\ee 
The table below shows the ratio of the dimensionless numbers 
$K$ and $h_+$ for various values of $f_+$: 
\vs{1} 

\begin{center}
\begin{tabular}{|c|c|}\hline  
$f_+$ (Hz) & $K/h_+$ \\ \hline 
 & \\
150 & $10^{-4}$ \\ 
 & \\
$1.5 \times 10^4$ & 1 \\ 
 & \\
$1.5 \times 10^6$ & $10^4$ \\ 
 & \\ 
$1.5 \times 10^9$ & $10^{10}$ \\ \hline
\end{tabular} 
\end{center} 

\nit
As $K$ grows quadratically with $f_+$, the ratio $K/h_+$ 
only becomes appreciable at high frequencies. At the 
frequencies at which bar antennae and interferometers 
are sensitive the perturbative influence of gravitational 
waves on particles in cyclotron orbit seems negligeable.  
 
Similarly, for bursts of gravitational radiation, which we 
model by an exponentially decaying amplitude 
\be 
\kgp\, =\, \frac{h_+}{2c^2 \tau^2}\, e^{-t/\tau}, \hspace{3em} 
  t > 0,
\label{3.12}
\ee 
the role of $\og_+$ is taken over by $1/\tau$. Thus 
appreciable ratios $K/
h_+$ are obtained only 
for very short bursts: $\tau \leq 10^{-5}$ sec, and 
preferably much less.  
\vs{1} 

\nit
Having discussed the variation in kinetic energy of particles 
under the influence of gravitational wave, we now consider the 
changes induced in the orbit itself. We start from the equations 
of motion (\ref{2.19}). To obtain an analytical result, we 
consider the special case of constant curvature components: 
$\kgp$, $\kgt$ =  constant. More complicated curvatures 
can be approximated by adding piece-wise constant curvature 
waves. The solution of the equations of motion than becomes 
\be 
x(u)\, =\, A\, \cos qu\, +\, B\, \sin qu, \hspace{2em} 
y(u)\, =\, \Lb\, \lh B \cos (qu - \ag) - A \sin (qu - \ag) \rh.
\label{3.13}
\ee 
Here the wavenumber $q$ can be expressed in terms of the 
cyclotron wavenumber $k_c$ and the curvature components as 
\be 
q^2\, =\, \frac{k_c^2}{2}\, \lh 1 + \sqrt{ \dsp{ 1 + 
  \frac{4|\kg|^2}{k_c^2}}} \rh\, \approx\, k_c^2\, +\, 
  \frac{|\kg|^2}{k_c^2}\, +\, ... 
\label{3.14}
\ee 
whilst the phaseshift $\ag$ and the scalefactor $\Lb$ are 
given by 
\be 
\ba{rll}
\tan \ag & = & \dsp{ \frac{\kgt}{qk_c}\, \approx\, 
  \frac{\kgt}{k_c^2}, } \\
 & & \\
\Lb & = & \dsp{ \frac{\sqrt{\kgt^2 + q^2 k_c^2}}{\kgp + q^2}\, 
  =\, \frac{qk_c}{\kgp + q^2}\, \frac{1}{\cos \ag}\, 
  \approx\, 1\, -\, \frac{\kgp}{k_c^2}\, +\, 
  \frac{\kgp^2}{2 k_c^4}\, +\, ... }
\ea 
\label{3.15}
\ee 
The solution (\ref{3.13}) describes an ellips defined by 
the orbital equation 
\be 
x^2\, -\, 2 \frac{\sin \ag}{\Lb}\, xy\, +\, \frac{y^2}{\Lb^2}\, 
  =\, (A^2 + B^2) \cos^2 \ag. 
\label{3.16}
\ee 
Rotating the co-ordinates by an angle $\fg$ such that 
\be 
\tan 2 \fg\, =\, \frac{2 \Lb \sin \ag}{\Lb^2 - 1}, 
\label{3.17}
\ee 
the equation can be cast in standard form in terms of the rotated 
co-ordinates $(\bar{x}, \bar{y})$ as 
\be 
\frac{\bar{x}^2}{\bar{a}^2}\, +\, \frac{\bar{y}^2}{\bar{b}^2}\, 
 =\, 1, 
\label{3.18}
\ee   
with 
\be 
\ba{rll}
\dsp{ \frac{1}{\bar{a}^2} }& = & \dsp{ \frac{1}{(A^2 + B^2) 
  \cos^2 \ag}\, \frac{\Lb^2 + 1}{2\Lb^2}\, \lh 1 + 
  \sqrt{\dsp{ 1 - \frac{4 \Lb^2 \cos^2 \ag}{(\Lb^2 + 1)^2} }} \rh, 
  } \\ 
 & & \\
\dsp{ \frac{1}{\bar{b}^2} }& = & \dsp{ \frac{1}{(A^2 + B^2) 
  \cos^2 \ag}\, \frac{\Lb^2 + 1}{2\Lb^2}\, \lh 1 - 
  \sqrt{\dsp{ 1 - \frac{4 \Lb^2 \cos^2 \ag}{(\Lb^2 + 1)^2} }} \rh, 
  } 
\ea 
\label{3.19}
\ee 
The eccentricity of the ellips is therefore given by 
\be 
e^2\, =\, \frac{\bar{b}^2 - \bar{a}^2}{\bar{b}^2}\, 
  \approx\, \frac{2 |\kg|}{k_c^2}. 
\label{3.20}
\ee 
For example, if $\kgt = 0$, we find the length of the minor 
and major semi-axes to be given by 

\be 
\bar{a}\, \approx\, \rg\, \lh 1 - \frac{\kgp}{k_c^2} \rh, 
 \hspace{2em} 
\bar{b}\, \approx\, \rg\, =\, \sqrt{A^2 + B^2},
\label{3.21}
\ee 
whereas in the case $\kgp = 0$ we obtain 
\be 
\bar{a}\, \approx\, \rg\, \lh 1 - \frac{\kgt}{2k_c^2} \rh, 
  \hspace{2em}  
\bar{b}\, \approx\, \rg\, \lh 1 + \frac{\kgt}{2k_c^2} \rh.
\label{3.22}
\ee
From eq.(\ref{3.3.1}) we recall that a particle moving at 
the speed of light in a near-circular orbit of radius $\rg$ 
has $k_c = 1/\rg$. Therefore the change in the orbital 
size is of the order of
\be 
\frac{\del \rg}{\rg}\, =\, \frac{1}{2}\, \rg^2\, |\kg|\, 
 =\, \sqrt{\langle K^2 \rangle}, 
\label{3.32}
\ee 
with the angular brackets denoting an average over a full 
orbit. Basically this shows that one may expect for the 
ratio $\del \rg/\rg$ numbers of the same order of magnitude 
as for $\del({\bf v}^2/c^2) \approx 2\, \del \og/\og_0$. 
In all cases an effect will be extremely difficult to 
establish, even at the highest gravitational-wave frequencies. 
However, as in general frequencies can be measured with higher 
precision than displacements, measurements of $\del \og/\og_0$ 
might be of interest to search for a high-frequency cosmic 
gravity-wave background.
\vs{2}

\nit
{\bf Acknowledgement} \nl 
The author gratefully acknowledges useful and stimulating discussions 
with Daniel Zer-Zion and Luc Vos of CERN, and Leonid Grischuk 
of Cardiff university.

\end{document}